\documentclass[letterpaper, 10 pt, conference]{ieeeconf}  

\IEEEoverridecommandlockouts                              

\usepackage[utf8]{inputenc}
\usepackage{authblk}
\usepackage{graphicx}
\usepackage{url}
\usepackage{listings}
\usepackage{xcolor}
\usepackage{amsmath}
\usepackage{comment}
\usepackage{hyperref}

\usepackage[colorinlistoftodos,prependcaption]{todonotes}
\definecolor{amber}{rgb}{1.0, 0.49, 0.0}
\definecolor{darkblue}{rgb}{0,0,0.5}
\definecolor{crimson}{rgb}{0.83, 0.0, 0.25}
\definecolor{darkgreen}{rgb}{0,0.5,0}
\definecolor{purple}{rgb}{0.57,0.63,0.81}
\definecolor{brown}{rgb}{0.65, 0.16, 0.16}
\definecolor{caribbeangreen}{rgb}{0.0, 0.8, 0.6}
\definecolor{carmine}{rgb}{0.59, 0.0, 0.09}
\definecolor{champagne}{rgb}{0.97, 0.91, 0.81}
\definecolor{classicrose}{rgb}{0.98, 0.8, 0.91}
\definecolor{corn}{rgb}{0.98, 0.93, 0.36}
\definecolor{cornflowerblue}{rgb}{0.39, 0.58, 0.93}
\definecolor{darkelectricblue}{rgb}{0.33, 0.41, 0.47}

\title{\LARGE \bf
Campaign Knowledge Network: Building Knowledge for  Campaign Efficiency
}

\author[1]{Sachith Withana}
\author[2] {Kshitij Mehta}
\author[2]{Matthew Wolf}
\author[1]{Beth Plale}
\affil[1]{Indiana University, Bloomington IN, USA}
\affil[2]{Oak Ridge National Laboratory, Oak Ridge TN, USA}

\begin{document}

\maketitle
\thispagestyle{empty}
\pagestyle{empty}

\begin{abstract}
In the landscape of exascale computing collaborative research campaigns are conducted as co-design activities of loosely coordinated experiments. But the higher level context and the knowledge of individual experimental activity is lost over time.  We undertook a knowledge capture and representation aid called Campaign Knowledge Network(CKN), a co-design design and analysis tool. We demonstrate that CKN can satisfy the Hoarde abstraction and can distill campaign context from runtime information thereby creating a knowledge resource upon which analysis tools can run to provide more efficient experimentation. 
\end{abstract}

\section{Motivation} \label{motiv}

The leap to exascale computing will bring about transformative progress in energy, life sciences, materials engineering, and national security. It will enhance our ability to tackle problems of a grand challenge scale.  Co-design has been promoted as a software development approach to address application development in an exascale setting. Co-design by definition is to design something jointly.  For exascale, this is a recognition that system parts, both hardware and software, must undergo design in synergy with and alongside other parts of the system.   

One specific problem for which the co-design approach is helpful is the gross imbalance between compute and output speeds. By 2024 computers are expected to compute at 1018 operations per second but write to disk only at 1012 bytes/sec, an extraordinarily high compute-to-output ratio that is 200 times worse than on the first petascale systems ~\cite{exscalesite}.  Co-design allows integrated design and testing of the rapidly developing software stack with emerging hardware technologies while developing software components that embody the most common application motifs~\cite{CODAR2020}.  Specific co-design exercises are called \textit{campaigns}: a multi-researcher co-design activity that contributes to a single scientific development objective \cite{mehta2019codesign}. 

Development and testing campaigns around the problem of high compute to output ratio, for instance, is conducted by teams of researchers who, among other activities, conduct large numbers of loosely coordinated performance experiments.  This loosely coordinated activity could benefit from an automated and intelligent observer that observes activity and distills this information into knowledge that augments the overall campaign objective.  

\textit{Campaign Knowledge Network (CKN)} is a co-design design and testing tool that constructs knowledge about experimental campaigns and provides analysis tools for decision making over the knowledge it constructs. Specifically, CKN is a graph-based knowledge network that captures and represents acquired knowledge about joint campaign activity.  It represents the campaign at a logical level and represents observational data gathered at runtime at a physical level. It distills physical level data to knowledge at the logical level. By the distillation process, CKN is designed to build and represent a complex, historical state graph for a campaign where the state graph represents a multidimensional parameter space, and clusters of points in the parameter space are where experimentation has occurred.    

Two forms of interaction with CKN are being developed as part of this research.  The first form, discussed and evaluated in this paper, is the interaction as established through the Hoarde abstraction ~\cite{logan2019vision}.  The second form is interaction through supported forms of analysis of CKN data. To this latter form, in this paper we evaluate techniques for similarity comparisons between experiments conducted in the same campaign. 

\begin{figure}[tbh]
  \includegraphics[width=\linewidth]{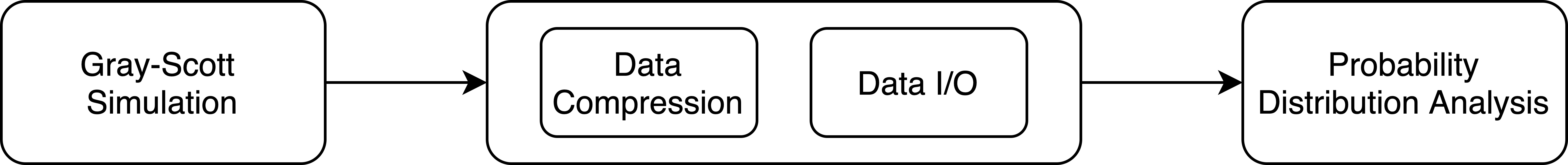}
  \caption{Gray-Scott reaction-diffusion model workflow}
  \label{fig:grayscott}
\end{figure}

Two applications provide experimental data for our analysis. The first is the the Gray-Scott reaction-diffusion model simulation \cite{pearson1993complex}. The model is paired with an analysis component that calculates the probability distribution of the simulation output, see Figure \ref{fig:grayscott}. 

The co-design experimentation is in the communication that occurs between the simulation and probability analysis, where the IO, orchestrated by ADIOS \cite{liu2014hello}, can select from a range of I/O paths and data compression techniques depending on a number of factors determined at run-time.   

The second application is the OSU Micro-benchmarks (MVAPICH) \cite{osumb}, specifically the point-to-point bandwidth test which calculates the communication bandwidth between two nodes in a HPC system.  

Cheetah and Savanna \cite{mehta2019codesign} provide the framework for experimental execution of sweep groups.  Cheetah is used to create execution plans for the experiment sweeps.  Savanna, on the other hand, deploys and manages these sweeps in the HPC systems. From a given campaign specification file, Cheetah creates each individual sweep and the sweep group corresponding to the specification. Savanna is invoked to run the individual run (instantiation of a sweep) in the given HPC system. Data transfer is via ADIOS, depicted in Figure \ref{fig:grayscott} as an explicit part of the workflow. 

The contribution of this paper is threefold: i) the conceptualization of the Campaign Knowledge Network (CKN); ii) an evaluation using CKN's early prototype system showing that it can satisfy the essential queries of the Hoarde abstraction; and iii) an evaluation of several techniques for comparing subgraphs in a co-design activity. We demonstrate that CKN is able to answer the queries for a majority of requirements imposed  by  the  Hoarde  abstraction  except  that  of requiring a  globally  scoped  PID  for  data  products  which  is ongoing work.

The remainder of this paper is organized as follows. Section \ref{sec:CKN} describes the Campaign Knowledge Network (CKN).    Section \ref{sec:CKNingest} describes the architecture of the prototype system. Section \ref{sec:hoarde} shows how CKN satisfies the essential queries of the Hoarde abstraction.  Section \ref{sec:similarity} evaluates similarity measures on subgraphs of CKN.  The paper concludes with related work in Section \ref{sec:relatedwork} and conclusion in Section \ref{sec:conclusion}.

\section{Campaign Knowledge Network (CKN)}\label{sec:CKN}

The Campaign Knowledge Network (CKN) is a knowledge graph representation of a campaign \cite{mehta2019codesign}.  A campaign is a scientific co-design development construct involving more than one hardware/software component.  As an example, a single campaign could be the exploration of which data I/O, data compression, and filtering techniques could be employed at any point in an application setting to reduce the effects of slow I/O speeds. A campaign captures the multi-party experimentation that is carried out within that larger discovery construct.  The campaign thus represents in both space and time the progression of discovery that occurs during a co-design.  CKN supports the campaign by  harvesting selective and strategic run time information and distills that as new knowledge about the state of the campaign through time.  CKN draws on data provenance for its data flow and time-based views of runtime activity.  

The objective of CKN is to aid scientific discovery through a knowledge representation and set of analysis tools that enable greater understanding of the complex state space of a multi-party campaign.  Through a campaign, multiple iterations of experiment sweeps are carried out to narrow down the parametric space. Depending on the experiment run time and the resource availability, this can take from days or months and involve multiple researchers.  

\begin{figure}
    \centering
    \includegraphics[width=\linewidth]{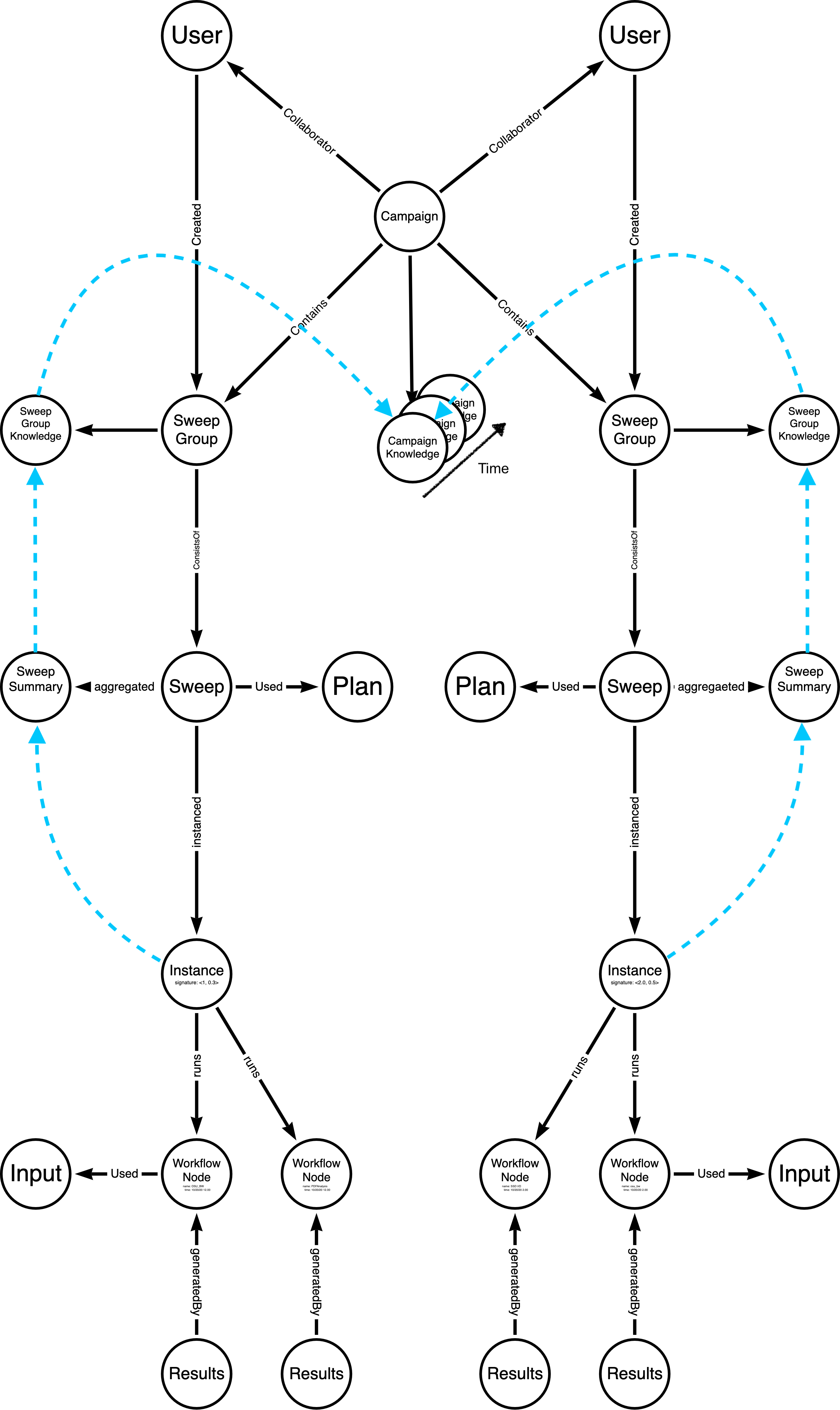}
    \caption{Knowledge Graph structure of a Campaign}
    \label{fig:campaignstructure}
\end{figure}

A backbone graph of CKN is captured in Figure \ref{fig:campaignstructure}. We adopt the terminology of Cheetah \cite{mehta2019codesign}, a tool for specifying co-design experiments.   Using this terminology, a scientific discovery process is a multi-user campaign that is made up of sweep groups consisting of sweeps as follows: 
\begin{itemize}
    \item 
\textit{A Campaign} is a multi-researcher co-design activity that contributes to a single scientific development objective. 
\item
\textit{Sweep group}. A campaign consists of multiple sweep groups, with each sweep group an experiment set up by a single researcher. A sweep group might, for instance, sweep across a couple parameters in the parameter space while holding the remaining parameters constant.
\item
\textit{A Sweep} is a member of a Sweep Group.  Sweeps are an individual experiment that is carried out with fixed parameter values. 
\item
\textit{Instance, Workflow Node} While campaigns, sweep groups, and sweeps are logical constructs, instances are physical or run-time constructs.  An Instance is an instantiation of a sweep, and is made up of one or more Workflow Nodes.  The workflow may be composed so that different executables reside on separate nodes, or share compute nodes, in addition to fine-tuning the number of processes per node.
\end{itemize}

A unique contribution of CKN is knowledge propagation of runtime information to the logical layer of the gaph to facilitate analysis.  The blue arrows of Figure \ref{fig:campaignstructure} (from Instance to Sweep Summary, from Sweep Summary to SweepGroupStatus, from SweepGroupStatus to CampaignStatus) capture this propagation of run time information from the physical layer through to the logical portions of the graph.   This distillation is synthesized and aggregated information about one or more run-time activities. In such a manner the upper levels of the graph represent historical experimental activity and aid  a computational scientist to mine past activity within a campaign to find similar experiments that were run. 

Specifically, Figure \ref{fig:campaignstructure} depicts a campaign with two collaborators, each working on their own separate sweep group. Each sweep group contains multiple sweeps with multiple instances for each as each sweep can be run multiple times. The subgraph starting with the instance node contains the physical manifestation of the experiment run where all the parameters, provenance and the results are stored. Everything above the instance node contains all the logical information related to the campaign. 


 

The prototype implementation of CKN uses the Komadu provenance capture system \cite{suriarachchi2015komadu} to ingest planning and performance data into Komadu in the form of data provenance \cite{simmhan2005survey} which derives process, actor, and entity information from low level performance monitoring information. The data provenance is filtered to represent both forward and backward provenance \cite{suriarachchi2018big}. CKN uses Neo4j to represent the graph and a distiller to distill data to knowledge for representation in the graph database.  CKN is a continuously evolving graph.

\section{Data Ingestion}\label{sec:CKNingest}

The campaign knowledge network is intended to augment the suite of planning and runtime tools (like Cheetah and Savanna) that support campaign experimentation.  In its prototype form, CKN is updated only after a sweep group has completed execution.   The higher latencies that this imposes for reacting and enacting change are acceptable for the science discovery level at which we are introducing automation; the approach will be examined in the future for more timely reaction times. 

A strength of CKN's approach to ingest is that data are transformed on the CKN side, thus lowering the burden of incorporating the solution into a different environment. 
CKN further does not require the data to be annotated in a certain way. Or data to be marked and stored differently the parallel file system hence providing the flexibility and re-usability of CKN with minimal user interference. Additionally, the information capture mechanism does not perturb experimentation since it piggybacks on existing logging tools and extracts information post-execution.  

\begin{figure}[tbh]
  \includegraphics[width=\linewidth]{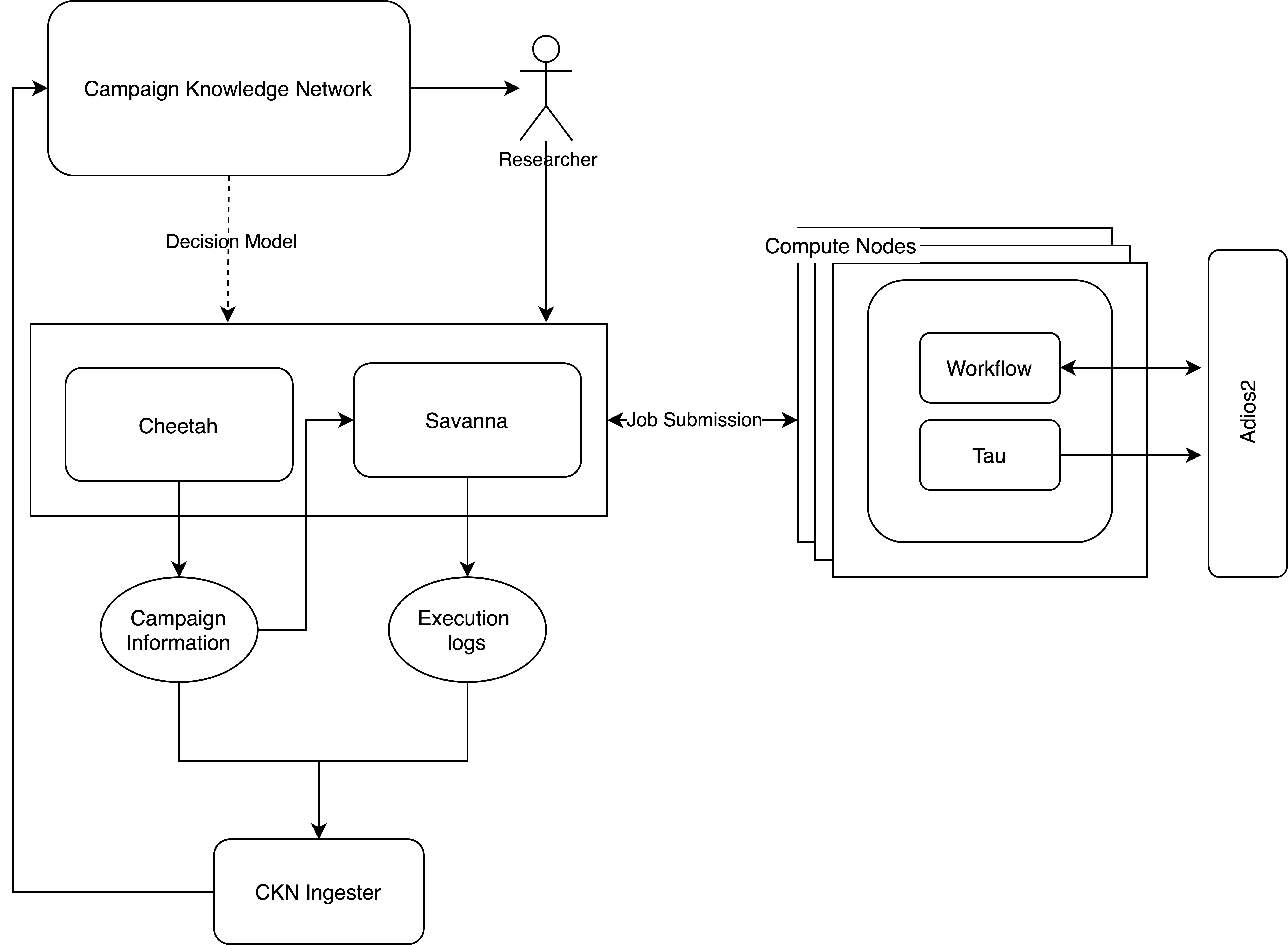}
    \caption{CKN in the context of codesign environment}
  \label{fig:cknintegration}
\end{figure}

The CKN data ingester is a portable set of scripts that extract the data from the system via log files and send the data to the CKN system. As depicted in Figure \ref{fig:cknintegration}, data are ingested into CKN from the Cheetah planning tool, and from various performance monitoring services that run in the HPC environment and are collected up  by Savanna.   The data ingester generates the graph nodes, relationships, and properties as run time provenance graphs.  A separate function distills aggregate information for updating the CKN campaign graph. 

Specifically in the context of working with Cheetah and Savannah, the CKN ingester is triggered post-run to extract information from logs written by Cheetah and Savanna. The information that Cheetah generates is transformed into \textit{prospective provenance} or plan-based provenance whereas the runtime information, such as ADIOS information obtained through Savannah is transformed into \textit{retrospective provenance} or runtime-based provenance \cite{lim2010prospective}.


The CKN Ingester supports two integration configuration methods for parsing the logs: polling and post-processing. 
\begin{itemize}
    \item Polling: where the Ingester monitors the campaign files and process the changes as they take place. 
    \item Post processing: where CKN ingester can be configured to run as a post processing stage in the workflow execution where it will then extract and process all the information available in the corresponding campaign logs. 
\end{itemize}

    


\section{Completeness with respect to Hoarde} \label{sec:hoarde}

One form of interaction with CKN is query based.  The other, through analysis tools provided as part of CKN, is future work and not discussed here.  The query interface is aligned with the Hoarde query abstraction  \cite{logan2019vision}. The Hoarde abstraction provides an abstract layer to manage large collections of data via their meta-data and provenance which makes the data products tracable for later use. Each atomic data item is treated as a Research Object (RO) \cite{bechhofer2010research} and are given unique id's for identification in the collection. This abstraction layer combined with the unique id's provides a comprehensive query interface to track and analyse these research object via queries such as i) finding datasets that satisfy a set of parameters, ii) finding the ancestors and descendants  of  a  dataset,  and iii) producing  the  sources  and products of a dataset ...etc.


CKN supports the set of essential queries identified by the Hoarde abstraction. It does so by utilizing the forward and backward  provenance  query  capabilities  that  CKN  inherits from  Komadu \cite{suriarachchi2018big}.  
Examples of each query is available in the Appendix. We use the Gray-Scott reaction diffusion model application that is run on Summit. These experiments were run using two nodes of Summit with multiple sweep groups with different campaigns. 

The primary queries in the Hoarde abstraction and CKN are as follows: 

\textbf{1. Dataset Discovery}
CKN satisfies this type of query through ID based lookup on the datasets. CKN additionally supports  search for a dataset based on key-value pair attributes, whereupon it returns all matching datasets for the given query. 

\textbf{2. Dataset Descendants and Ancestors}
It is crucial to understand and query the life cycles of the datasets we have available. When starting a new experiment, one needs to be able to query and understand what experiments were run and what data products were produced in order to narrow down the search space for the experiment bootstrapping. Hence we employ the forward provenance queries available in CKN for this specific purpose, to find the descendants of datasets. 

Similarly, using backward provenance queries, we are able to find the Ancestors of a given dataset. This enables the users to further explore the search space to evaluate the different combinations used to produce the output products of experiments. 

CKN draws on Komadu \cite{suriarachchi2018big} for support for forward and backward provenance queries. The combination of these two powerful tools provides the user the ability to explore the CKN in terms of data products to further the understanding of the available CKN experiments. 

\textbf{3. Dataset Sources and Products}
Once narrowed down, the user needs to understand how the data product was produced and what other uses were for that dataset. What other workflows used the same dataset? and how? These are the sources and products queries which directly corresponds to workflow provenance queries. 

To understand the sources of a given dataset, to analyse how a particular data product is produced, CKN provides the complete provenance graph for that data product as the source (either comprehensive or concise based on required detail level). It shows the workflow information resulting on how the data product was produced. The same is applied to find the products of a dataset, using a complete provenance query in CKN. 

Building off of the aforementioned queries, CKN allows the user to narrow down the datasets and the workflows (experiment sweeps) the user is interested in. It allows the user to understand individual workflows and their purpose.

The unique addition of the Knowledge Network for campaign data extends these queries beyond the scope of traditional data provenance queries. Via the Knowledge Network, CKN allows the user to step back and explore beyond few individual workflows into the realm of campaigns.

\section{Similarity Comparison} \label{sec:similarity}

Experimentation carried out in the context of a campaign is carried out within a state space of hyperparameters.   Any individual experiment is thus a somewhere in this multi-dimensional space.   If CKN can help a researcher understand where their experimentation sits within this complex state space especially relative to a colleague's experimentation, then time can be saved and efficiencies had.  For CKN, this problem is one of detecting similarity between subgraphs.

We evaluate several different techniques for detecting similarity between subgraphs.  Our approach identifies a feature vectors for each experiments, computes a signature for each experiment on its feature vector, and defines a similarity measure as a distance between signatures.  This approach draws from similarity based clustering that is done in data mining \cite{huang2008similarity}.

Data for use in this study is collected by using data from the Gray-Scott reaction diffusion model experiment discussed in Section \ref{sec:hoarde}.  We further conduct a campaign by two users on two different HPC systems (Summit at ORNL and BigRed3 at Indiana University) using the Ohio State University Micro Benchmark(OSU MB) \cite{osumb}. Each OSU experiment was run using two nodes (one per application) on both Summit and BigRed3 to measure the point-to-point communication bandwidth between these nodes. 

Experiment signatures have two primary benefits.  The first is re-usability of the exactly similar experiment data. With the large compute times and cost of the current scientific workflows, it is critical to reduce the redundancies encountered via the workflow system. Therefore CKN provides a way to first compare if an experiment with the provided parameters have already been run in the system and if so, what were the results. This added level of abstraction provides the user the ability to determine if that particular experiment needs to be re-run or if the data from the previous experiments can be re-used saving invaluable compute times. If the exact similar experiments had any issues (failed or taken longer than required), it allows the user to adjust their parameters and resource allocations to rectify these issues saving another compute run on the workflow. Going forward we would be evaluating the option of adding the ability to make these decisions into the workflow system (Cheetah and Savanna) itself so that once the campaign is generated, the re-used data can be populated as well. 

 The second benefit of using the similarity metric is that the researchers do not have to run all the nodes in the parametric space. They can employ the signatures to explore the experiments \textit{similar} to a given set of parameters. It allows the convenience of evaluating the results of the tree and similar experiments without actually having to run those experiments. The idea is to influence informed decision making via similarity measures in CKN to explore and re-use the information in this large system for pure efficiency. 

\subsection{Signature definition}
There are numerous types of workflows being run with different input parameters and types. Hence the signature depends on those different parameters to precisely compare \textit{similar} workflow experiments which made the feature selection workflow specific. Each workflow type would have its own type of feature vector due to the dissimilarities between different workflows. We describe the feature extraction for the Gray-Scott model. 

Gray-Scott reaction diffusion model is based on the chemical reactions of chemicals U and V producing P according to the following two equations. 
\[
\begin{aligned}
U+2 V & \rightarrow 3 V \\
V & \rightarrow P
\end{aligned} 
\]

This reaction is simulated via solving these partial differential equations:
\[ 
\begin{aligned}
\frac{\partial u}{\partial t} &=d_{u} \nabla^{2} u-u v^{2}+f(1-u) \\
\frac{\partial v}{\partial t} &=d_{v} \nabla^{2} v+u v^{2}-(f+k) v
\end{aligned}
\]

Here u, v are the the chemical concentrations, f is the feed rate, k is the conversion rate from V to P, du and dv are diffusion rates. Apart from these parameters, the simulation grid size(L) is also added to the model. Gray-Scott is highly non-linear, hence a small change in the parameters varies the results drastically. Therefore we needed to capture this non-linearity in the captured feature space to represent a single simulation. Then a feature vector can be constructed via these parameters that would take the form:
\[
<\mathrm{L}, \mathrm{D} \mathrm{u}, \mathrm{D} v, \mathrm{~F}, \mathrm{k}>
\]

Each vector space is fine as it is (needs no further refinement). We explain it as a working set of values rather than a complete vector space.

\subsection{Signature Comparison}


 This feature vector is used as the \textit{signature} for a Gray-Scott simulation experiment. This signature can easily be extended to include different parameters such as ADIOS \cite{liu2014hello} configurations and machine information. This signature is calculated at the CKN ingester and injects it into the Knowledge graph for the similarity measure querying. The signature is added to the instance node of the graph for the comparisons. 

A similarity measure is a distance measure between experiments (sweeps), see Figure \ref{fig:campaignstructure}, dotted  line  between  sweeps.   
One approach to computing similarity is to compare two vector signatures and quantify the deviation in terms of a singular value if they are not the same. The main feature we looked for such a measure was that it has to be symmetric, therefore the similarity from X to Y would be the same as from Y to X. Hence, inspired by other literature in the machine learning space \cite{nguyen2010cosine, li2011measuring}, we evaluated three similarity measures, Cosine similarity, Euclidean and Manhattan distance based similarities.

\begin{figure}
  \includegraphics[width=\linewidth]{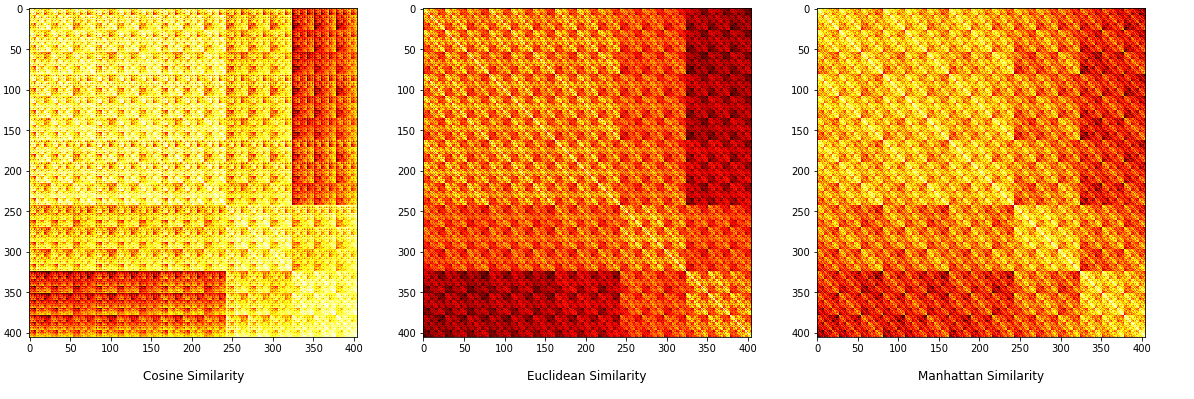} 
  \caption{Similarity Metrics Comparison: The figure is comparing each experiment run to all other runs. Darker the color, lower the similarity values between the signatures.}
  \label{fig:simcomp}
\end{figure}

To evaluate these metrics we take a real-world Gray-Scott Campaign and create the vector signatures for each possible signature. Then we evaluate and compare the similarity values obtained using the aforementioned metrics. The results are shown in Figure \ref{fig:simcomp}. 

Using the cosine metrics, the values obtained are with very small variance. Therefore the whole spectrum of similarity values are compressed into few decimal points instead of the whole range. We noticed that as the number of parameters grow in the feature vector, due to the small ranges of each parameters, the cosine metrics perform worst. The cosine similarity values spread can be seen in the 2D heatmap on the left in the Figure. 

But in the distance metrics (Euclidean and Manhattan distance), the results are much more spread out and provides a higher range of values compared to the cosine metrics which makes the comparisons much more viable. As seen in the 2D Heatmap to the right in the figure, the spread of the values are more predictable and clear as opposed to the cosine metrics. The resulting 2D Heatmap obtained for Manhattan distance is exactly similar to the one depicted for the Euclidean metrics. Therefore we decided to use the Euclidean Similarity metric instead of the Cosine metric for the signature comparisons in CKN.


\subsection{Discussion}
\paragraph{Similarity Measure use in CKN}
A campaign is a continuously evolving multi-party effort. The researchers continue exploring the interested parametric and the configuration space of the research until a solution is found. During the life-cycle of these campaigns involving multiple researchers, the status of the whole campaign might not be clearly visible to all as these experiments are run on multiple HPC systems with their own parametric explorations. 

Since the experiments are conducted by different users on different systems, to understand the bigger picture of a campaign, an accumulation and analysis of all the experiments are required. Therefore we evaluate methods to aggregate sweep information towards the campaign itself by propagating information up through the graph.

The benefits of this is two fold. First it allows the researchers to evaluate the status of the campaign. To answer questions like how complete or not the campaign effort is. Secondly, it allows researchers to take informed decisions on further exploration of significant sweep groups by extending them. 

Therefore after the completion of each sweep group, an aggregation is run along all the experiments to accumulate and summarize the parameters and the results obtained in the sweep group. This information is then propagated up towards the Campaign node in the Knowledge Graph to create the "Campaign Status" for that particular campaign. This is depicted in Figure \ref{fig:campaignstructure} using blue dashed lines. Starting with the instance summary, it's propagated towards the sweep group and then onto the Campaign status node. In future work, the comparison matrix for all the experiments run would be added to the status nodes of the aforementioned knowledge graph.



As described in Section \ref{sec:CKN} the information contained in the subgraph for each experiment are summarized, aggregated and propagated towards the campaign node to create the campaign status.


\section{Related Work} \label{sec:relatedwork}

Related to our research is Performance improvements through provenance and visualization.  Pouchard et al. \cite{pouchard2017capturing} demonstrates provenance-based performance optimizations and evaluations of scientific workflows on extreme scale HPC systems. Combining a provenance tool (ProvEn) and a profiling tool (TAU \cite{shende2006tau}), the resulting Chimbuko framework combines both provenance and profiling and tracing aspects of workflows together for better evaluation via visualizing. In \cite{pouchard2018prescriptive}, the authors improve upon the framework to control the volume of data stored via in situ anomaly detection. Using ADIOS \cite{liu2014hello} as the in situ I/O library, it detects anomalies and stores the performance data only if an anomaly is detected. They define the term "Prescriptive provenance" which includes both the traditional provenance lineage data (static code, metadata) and the provenance of the execution of the scientific workflows which includes runtime environment. software stack, system information ...etc.


Monitoring Analytics Infrastructure (MONA) provides in situ performance analysis for high performance computing workflows and coupled codes through ADIOS2. Using a data aggregation service for HPC named Scalable Observation System (SOS)\cite{wood2016-sosflow}, the performance metrics extracted from existing tools such as TAU are aggregated throughout the distributed workflows. The authors present different use cases for in situ performance analytics such as performance monitoring and communication pattern monitoring and evaluates the trade off between exhaustively capturing all performance information vs not capturing enough for feature recognition.

Reproducibility of workflows depends not only on the provenance of the workflows but also on the underlying infrastructure.  Santana-Perez et al. \cite{santana2017reproducibility} uses logical preservation of the infrastructure by conserving the descriptions of the execution environment. This is achieved via a semantic modeling to describe the resources involved in the environment using semantic vocabularies. 

The efficient capture and representation of campaign knowledge could benefit the campaign management tools by allowing recommendations. CKN as the basis for an AI decision model is an objective. Baracaldo et al. \cite{baracaldo2017mitigating} use AI on captured provenance to detect poisonous data in adversarial environments. To avoid feeding  poisonous data into their machine learning algorithm, the authors employ an AI based algorithm to compare classification accuracies and remove the poisonous data based on the lineage of the data.

Komadu\cite{suriarachchi2015komadu} is a provenance repository built supporting the W3C prov standard\cite{missier2013w3c} for data provenance.  We employ the use of Komadu \cite{suriarachchi2015komadu}, predecessor to Karma \cite{simmhan2008karma2} as the provenance repository used in the Campaign Knowledge Network. Being decoupled from any specific scientific workflows allows Komadu to adapt freely to any incoming use cases with no modifications. The use of event-driven data ingestion allows Komadu to be completely independent from any workflow managers and the use of W3C Prov standard \cite{missier2013w3c} ensures the portability and reusability of the provenance captured. 


\section{Conclusion} \label{sec:conclusion}

The Campaign Knowledge Network represents information about campaigns that could be used to make experimentation more efficient and to automate the design of new experiments.  The proof of concept exploratory research described here provides early favorable results.   By leveraging the existing Komadu system, CKN is responsive to the Hoarde API. In tests on a reaction-diffusion model simulation we are able to obtain early favorable results for finding similarity between two sweeps. Associating results with persistent IDs is ongoing work.     

Ongoing work is to further develop the contextual information about the campaign, including capturing change through time and transitions of data (such as a model representation of a dataset) and across different researchers within a campaign.  Rich contextual information can be mined for automating workflow planning.   In this case, CKN closes the loop between plan information, run time results, and suggested optimizations for future workflow planning. 

CKN would influence both static and dynamic decision making systems for efficient experiment executions. It would help provide decisions such as what compression strategy to use, what scheduling parameters (number of nodes, node layout ...etc) to use, time of day scheduling ...etc. CKN would also provide debugging of the workflows to figure out the ramifications and their solutions. Furthermore, we would be adding a real-time workflow monitoring system via CKN using both real-time provenance and performance information through Tau and ADIOS2. 

A scientific workflow would involve multiple datasets. Therefore we recognize the importance of supporting the Hoarde queries for a set of data elements treating them as a single entity. Therefore it would allow queries such as forward provenance queries on a set of datasets instead of a single data entity. 

\section{Acknolwedgements}
This work funded in part by award \#1234983 by the National Science Foundation and in part by the Lilly Endowment, Inc., through its support for the Indiana University Pervasive Technology Institute. We offer our sincere gratitude to the Oak Ridge National Laboratory for the internship opportunity through which the idea of CKN emerged.  

\bibliography{main}
\bibliographystyle{plain}

\clearpage
\appendix

Find dataset:
\begin{lstlisting}[language=XML]
<quer:findActivityRequest
    xmlns:quer="http://komadu.d2i.indiana.edu/query"
    xmlns:komadu="http://komadu.d2i.indiana.edu">
    <quer:name>grayscott</quer:name>
    <quer:attributeList>
        <komadu:attribute>
            <komadu:property>U</komadu:property>
            <komadu:value>10</komadu:value>
        </komadu:attribute>
        <komadu:attribute>
            <komadu:property>L</komadu:property>
            <komadu:value>4</komadu:value>
        </komadu:attribute>
    </quer:attributeList>
</quer:findActivityRequest>
\end{lstlisting}

List the lineage of a dataset:
\begin{lstlisting}[language=XML]
<quer:getEntityGraphRequest 
xmlns:quer="http://komadu.d2i.indiana.edu/query"
xmlns:komadu="http://komadu.d2i.indiana.edu">
  <quer:entityType>FILE</quer:entityType>
  <quer:entityURI>swithana-ftest-run-1.iteration-0-simulation-stdout</quer:entityURI>
  <quer:informationDetailLevel>FINE</quer:informationDetailLevel>
</quer:getEntityGraphRequest>
\end{lstlisting}

Sources of a dataset:
\begin{lstlisting}[language=XML]
<quer:getEntityBackwardGraphRequest
xmlns:quer="http://komadu.d2i.indiana.edu/query"
xmlns:komadu="http://komadu.d2i.indiana.edu">
  <quer:entityType>FILE</quer:entityType>
  <quer:entityURI>swithana-kTest-ftest-run-0.iteration-0-simulation-stderr</quer:entityURI>
  <quer:informationDetailLevel>FINE</quer:informationDetailLevel>
</quer:getEntityBackwardGraphRequest>
\end{lstlisting}

Products of a dataset:
\begin{lstlisting}[language=XML]
<quer:getEntityForwardGraphRequest
xmlns:quer="http://komadu.d2i.indiana.edu/query"
xmlns:komadu="http://komadu.d2i.indiana.edu">
  <quer:entityType>FILE</quer:entityType>
  <quer:entityURI>swithana-kTest-ftest-run-0.iteration-0-settings.json</quer:entityURI>
</quer:getEntityForwardGraphRequest>
\end{lstlisting}

\end{document}